# Continuous-Aperture Array Based OAM High-Capacity Communication For Metaverse

Hongyun Jin[†], Wenchi Cheng[†], Jingqing Wang[†], and Wei Zhang[‡]

[†]State Key Laboratory of Integrated Services Networks, Xidian University, Xi'an, China
[‡]School of Electrical Engineering and Telecommunications, the University of New South Wales, Sydney, Australia
E-mail: {*hongyunjin@stu.xidian.edu.cn, wccheng@xidian.edu.cn, jqwangxd@xidian.edu.cn*, and *w.zhang@unsw.edu.au*}

*Abstract*—The extensive data interaction demands of an immersive metaverse necessitate the adoption of emerging technologies to enable high-capacity communication. Vortex electromagnetic waves with different orbital angular momentum (OAM) modes are spatially orthogonal, providing a novel spatial multiplexing dimension to achieve high-capacity communication. However, the number of orthogonal OAM modes based on a discrete uniform circular array (UCA) is limited by the number of array-elements in the UCA, and traditional discrete channel models are unable to accurately capture the physical properties of vortex electromagnetic wave propagation. The continuous-aperture array (CAPA) is composed of densely packed electromagnetic excitation elements, capable of flexibly and efficiently generating the desired surface currents to produce an arbitrary number of mutually orthogonal OAM modes. From the perspective of electromagnetic information theory (EIT), we propose a CAPA-based OAM orthogonal transmission scheme to realize high-capacity communication. We design the surface currents of the CAPA using Fourier basis functions, derive the electromagnetic channel for vortex electromagnetic waves, and investigate the upper bound of the spectrum efficiency for CAPA-based OAM orthogonal transmission. This paper establishes a theoretical foundation for applying EIT to the orthogonal transmission of vortex electromagnetic waves, offering a novel solution for achieving CAPA-based efficient and high-capacity communication.

## I. Introduction

THE Metaverse is a parallel world closely connected to the real world, arising from the development and integration of multiple technologies and representing the next stage in the evolution of the Internet [1]. The extensive data interaction demands of immersive experiences in the Metaverse necessitate emerging technologies to enable massive communication [2]. Vortex electromagnetic waves carrying orbital angular momentum (OAM) have attracted extensive research attention as an emerging technology in wireless communications [3]. The vortex electromagnetic waves with different OAM modes are spatially orthogonal, providing a novel spatial multiplexing approach for massive communications [4].

The continuous-aperture array (CAPA) is realized through continuous surfaces that integrate subwavelength-spaced antennas or metamaterials. Such surfaces, composed of densely packed electromagnetic (EM) excited elements, can manipulate impinging fields with remarkable flexibility and precision, thereby efficiently generating arbitrarily desired electromagnetic waves [5]. The authors of [6], [7] utilized the eigenfunctions of appropriate operators to represent the spatially continuous transmit currents and receive fields, thereby determining the capacity bound of continuous-space electromagnetic channels. The authors of [8] proposed a generalized EM-domain channel model and analyzed the capacity limits of systems with arbitrarily placed surfaces. The limited number of array-elements in traditional discrete uniform circular arrays (UCAs) constrains the number of mutually orthogonal OAM modes they can generate, thereby limiting the performance of UCA based OAM orthogonal transmission [9]. In contrast, CAPA based OAM orthogonal transmission enables the generation of an arbitrary number of OAM modes and pure vortex electromagnetic waves, making it highly suitable for high-capacity communication.

Electromagnetic information theory (EIT) serves as a unified framework that integrates electromagnetic theory, information theory, antenna theory, and wireless propagation channel modeling theory [10]. As an interdisciplinary field, EIT combines deterministic EM theory and statistical information theory to investigate information transmission in continuous EM fields [11]. The authors of [12] demonstrate how EIT provides a physics-aware information theory for analyzing and enhancing contemporary communication models. Additionally, they investigate critical issues such as subwavelength antenna design, electromagnetic mutual coupling, and spatial multiplexing in near-field channels from the perspective of EIT. The authors of [13] introduce an EIT based model that efficiently characterizes MIMO systems in complex environments, surpassing the limitations of traditional free-space approaches. However, current studies on EIT seldom address the specific design of orthogonal basis functions, instead focusing on broad and generalized frameworks. The different OAM modes of vortex electromagnetic waves are inherently based on Fourier orthogonal basis functions. Thus, investigating CAPA based OAM orthogonal transmission from the perspective of EIT is both essential and unavoidable.

In order to solve the above-mentioned problems, we propose the CAPA based OAM orthogonal transmission scheme to enhance the information-carrying capacity of electromagnetic waves, while ensuring that electromagnetic waves carrying different symbols remain orthogonal at the receiver, thereby enabling high-capacity communication. We investigate the coupling strength and determine the maximum number of orthogonal data streams that can be effectively transmitted between the transmitter and receiver regions, even though the CAPA based OAM orthogonal transmission can theoretically generate an infinite number of mutually orthogonal OAM modes. Furthermore, we explore the upper bound of the spectrum efficiency (SE) for CAPA based OAM orthogonal transmission. Exploring vortex electromagnetic waves from the perspective of EIT enables a more effective integration of vortex electromagnetic waves with information-carrying orthogonal basis functions. From the perspective of EIT, we design surface currents using Fourier basis functions to generate vortex electromagnetic waves, thereby transforming the abstract and generalized orthogonal transmission framework into a practical and implementable spatial orthogonal transmission scheme.

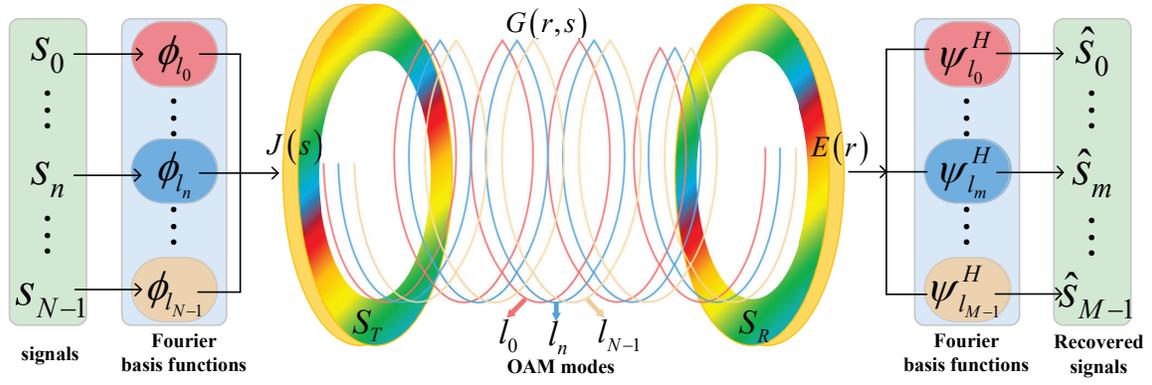

Fig. 1. The system model for CAPA based OAM orthogonal transmission.

The rest of this paper is organized as follows. In Section II, the system model for CAPA based OAM orthogonal transmission are provided. In Section III, the transmission schemes and the CAPA based electromagnetic channel model are analyzed. Section IV provides the numerical results. Finally, we conclude this paper in Section V.

**Notation**: Matrices and vectors are denoted by the capital letters and the lowercase letters in bold, respectively. For a given vector $\mathbf{p}$, $\hat{\mathbf{p}}$ is a unit vector along its direction and $\|\mathbf{p}\|$ denotes its magnitude. The notations $(\cdot)^H$ denote the Hermitian of a matrix or a vector. The curl operator $\nabla \times$ measures the rotation of a field, while the divergence operator $\nabla \cdot$ measures the total flux exuding from a point.

## II. THE SYSTEM MODEL FOR CAPA BASED OAM ORTHOGONAL TRANSMISSION

Figure 1 depicts the system model for CAPA based OAM orthogonal transmission. At the transmitter, the orthogonal basis function $\phi_{l_n}$ ($n \in [0, N-1]$) is used for the modulation of OAM mode $l_n$, with $N$ orthogonal basis functions being mutually orthogonal. The $N$ signals $s_n$ ($n \in [0, N-1]$) are respectively loaded onto the $N$ orthogonal basis functions $\phi_{l_n}$ for orthogonal transmission. Theoretically, a CAPA is capable of transmitting an arbitrary number of OAM modes. In this paper, the CAPA is modeled as a circular ring of homogeneous medium, with the CAPA transmitter and receiver regions denoted as $S_T$ and $S_R$, respectively. In the ideal case, the CAPA features a continuous antenna aperture, capable of generating any current distribution $\boldsymbol{J}(\boldsymbol{s}) \in \mathbb{C}^{3 \times 1}$ on its continuous surface. The CAPA composed of densely packed electromagnetic excited elements, can manipulate impinging fields with remarkable flexibility and precision, thereby efficiently generating arbitrarily desired electromagnetic waves.

Here, we focus on the scenario where the CAPAs at both the transmitter and receiver are aligned. The vortex electromagnetic waves carrying $N$ OAM modes are transmitted by the transmitter CAPA and received by the receiver CAPA. The dyadic Green's function $\boldsymbol{G}(\boldsymbol{r}, \boldsymbol{s}) \in \mathbb{C}^{3 \times 3}$ in free space is used to describe the interaction between the current $\boldsymbol{J}(\boldsymbol{s})$ at point $\boldsymbol{s} \in S_T$ and the electric field $\boldsymbol{E}(\boldsymbol{r}) \in \mathbb{C}^{3 \times 1}$ at point $\boldsymbol{r} \in S_R$. At the receiver, the orthogonal basis function $\psi_{l_m}$ ($m \in [0, M-1]$) is used for the demodulation of OAM mode $l_m$, and the $M$ orthogonal basis functions are mutually orthogonal. The signal $\hat{s}_m$ carried by OAM mode $l_m$ is recovered after OAM demodulation, leveraging the orthogonality among the orthogonal basis functions.

## III. THE CAPA BASED OAM TRANSMISSION SCHEME

### A. The dyadic Green's function

The Helmholtz wave equation in a homogeneous, isotropic medium is given by:

$$\nabla \times \nabla \times \boldsymbol{E}(\boldsymbol{r}) - k_0^2 \boldsymbol{E}(\boldsymbol{r}) = i\omega\mu \boldsymbol{J}(\boldsymbol{r}) = ik_0 Z \boldsymbol{J}(\boldsymbol{r}), \quad (1)$$

where $k_0 = \frac{w}{c} = \frac{2\pi}{\lambda}$ is the wavenumber in free space at angular frequency $\omega$, $\lambda$ is the wavelength, $Z = \sqrt{\frac{\mu}{\varepsilon}}$ is the impedance of free-space, and $\mu$, $\varepsilon$ are the permittivity and permeability, respectively. We denote $\boldsymbol{E}(\boldsymbol{r})$ as the electric field (in $volts/m$) at point $\boldsymbol{r}$, and $\boldsymbol{J}(\boldsymbol{r})$ as the current density (in $amperes/m^2$) at point $\boldsymbol{r}$.

Then, with $\nabla \times \nabla \times \boldsymbol{E}(\boldsymbol{r}) = -\nabla^2 \boldsymbol{E}(\boldsymbol{r}) + \nabla \nabla \boldsymbol{E}(\boldsymbol{r})$ and $\nabla \cdot \boldsymbol{E}(\boldsymbol{r}) = \nabla \cdot \boldsymbol{J}(\boldsymbol{r})/i\omega\varepsilon$, we further rewrite Eq. (1) as:

$$\nabla^2 \boldsymbol{E}(\boldsymbol{r}) + k_0^2 \boldsymbol{E}(\boldsymbol{r}) = -i\omega\mu \left[ \bar{\boldsymbol{I}} + \frac{\nabla\nabla}{k_0^2} \right] \cdot \boldsymbol{J}(\boldsymbol{r}), \quad (2)$$

where $\bar{\boldsymbol{I}}$ is an identity operator.

The Green's function of a wave equation represents the solution corresponding to a point source. Once the solution for a point source is obtained, the solution for a general source can be derived using the principle of linear superposition. This follows from the linearity of the wave equation, as any general source can be expressed as a linear superposition of point sources.

In this work, we adopt a continuous approach to study the CAPA based OAM transmission, as discrete formulations may fail to accurately capture fundamental physical properties of electromagnetic wave propagation. The process of electromagnetic wave propagation can be viewed as a linear system, with the dyadic Green's function acting as its impulse response. The Green's function is the solution to the wave equation Eq. (2). By introducing Green's function, the electric field $\boldsymbol{E}(\boldsymbol{r})$ can be given as:

$$\begin{aligned} \boldsymbol{E}(\boldsymbol{r}) &= i\omega\mu \int_{S_T} g(\boldsymbol{r},\boldsymbol{s}) \left[\bar{\boldsymbol{I}} + \frac{\nabla\nabla}{k_0^2}\right] \boldsymbol{J}(\boldsymbol{s}) d\boldsymbol{s} \\ &= i\omega\mu \int_{S_T} \boldsymbol{G}(\boldsymbol{r},\boldsymbol{s}) \boldsymbol{J}(\boldsymbol{s}) d\boldsymbol{s}, \end{aligned} \quad (3)$$

where $g(\boldsymbol{r},\boldsymbol{s})$ is the unbounded medium scalar Green's function and can be given as follows:

$$g(\boldsymbol{r},\boldsymbol{s}) = g(\boldsymbol{s},\boldsymbol{r}) = g(\boldsymbol{s}-\boldsymbol{r}) = \frac{e^{ik_0\|\boldsymbol{r}-\boldsymbol{s}\|}}{4\pi\|\boldsymbol{r}-\boldsymbol{s}\|}. \quad (4)$$

The dyadic Green's function $\boldsymbol{G}(\boldsymbol{r},\boldsymbol{s}) \in \mathbb{C}^{3\times 3}$ relates the vector electric field $\boldsymbol{E}(\boldsymbol{r}) \in \mathbb{C}^{3\times 1}$ to the vector current source $\boldsymbol{J}(\boldsymbol{s}) \in \mathbb{C}^{3\times 1}$, can be given as follows:

$$\begin{aligned} \boldsymbol{G}(\boldsymbol{r},\boldsymbol{s}) &= \left[\bar{\boldsymbol{I}} + \frac{\nabla\nabla}{k_0^2}\right] g(\boldsymbol{r},\boldsymbol{s}) = \frac{ik_0 Z e^{ik_0\|\mathbf{p}\|}}{4\pi \|\mathbf{p}\|} \cdot \\ &\left[\bar{\boldsymbol{I}} - \hat{\mathbf{p}}\hat{\mathbf{p}}^H + \frac{i\lambda}{2\pi\|\mathbf{p}\|}\left(\bar{\boldsymbol{I}} - 3\hat{\mathbf{p}}\hat{\mathbf{p}}^H\right) - \frac{\lambda^2}{(2\pi\|\mathbf{p}\|)^2}\left(\bar{\boldsymbol{I}} - 3\hat{\mathbf{p}}\hat{\mathbf{p}}^H\right)\right], \end{aligned} \quad (5)$$

where $\hat{\mathbf{p}} = \mathbf{p}/\|\mathbf{p}\|$ denotes the unit direction vector, and $\mathbf{p} = \boldsymbol{r} - \boldsymbol{s}$ denotes the distance vector.

### B. The Orthogonal Basis Functions Of OAM Modes

The current density is a sinusoidal function of time, which means it is time-harmonic. It can be expressed as:

$$\boldsymbol{J}(\boldsymbol{s},t) = \mathfrak{Re}\left[\boldsymbol{J}(\boldsymbol{s}) e^{-i\omega t}\right]. \quad (6)$$

This representation allows us to disregard the time-dependent component $e^{-i\omega t}$ and focus solely on the time-independent current density $\boldsymbol{J}(\boldsymbol{s})$.

The orthogonal basis functions of OAM modes simplify the implementation of the communication system and provide a clear physical interpretation. The current density $\boldsymbol{J}(\boldsymbol{s})$ can be expressed as a series of orthogonal basis functions $\phi_{l_n}$ ($n \in [0, N-1]$) [14] in the source space:

$$\boldsymbol{J}(\boldsymbol{s}) = \sum_{n=0}^{N-1} \xi_n \phi_{l_n}. \quad (7)$$

Here, the coefficient $\xi_n$ represents the weight of each basis function and is given by:

$$\xi_n = \int_{S_T} \phi_{l_n}^H \boldsymbol{J}(\boldsymbol{s}) d\boldsymbol{s}, \quad (8)$$

where this formulation is derived using Fourier basis functions to represent the spatially continuous transmitter current.

The orthogonal basis function $\phi_{l_n}$ of OAM mode $l_n$, can be given as follows [15]:

$$\phi_{l_n} = \frac{e^{il_n\varphi}\hat{\boldsymbol{j}}}{\sqrt{2\pi R_s}}, \quad (9)$$

where $\hat{\boldsymbol{j}}$, $R_s$, and $\varphi \in [0, 2\pi]$ represent the unit vector, the transmitter CAPA radius, and the azimuth angle within the transmitter CAPA, respectively.

Based on the orthogonality between any two basis functions, the orthogonal basis functions $\phi_{l_n}$ satisfy:

$$\int_{S_T} \phi_{l_n} \phi_{l_{n'}}^H ds = \begin{cases} \|\phi_{l_n}\|, l_n = l_{n'}; \\ 0, otherwise. \end{cases} \quad (10)$$

In the receiver region $S_R$, the electric field $\boldsymbol{E}(\boldsymbol{r})$ can be expressed as a series of orthogonal basis functions $\psi_{l_m}$ ($m \in [0, M-1]$):

$$\boldsymbol{E}(\boldsymbol{r}) = \sum_{m=0}^{M-1} \alpha_m \psi_{l_m}. \quad (11)$$

The coefficient $\alpha_m$ represents the weight of each basis function and satisfies:

$$\alpha_m = \int_{S_R} \psi_{l_m}^H \boldsymbol{E}(\boldsymbol{r}) d\boldsymbol{r}. \quad (12)$$

The orthogonal basis function $\psi_{l_m}$ of OAM mode $l_m$, can be given as follows:

$$\psi_{l_m} = \frac{e^{il_m\theta}\hat{\boldsymbol{e}}}{\sqrt{2\pi R_r}}, \quad (13)$$

where $\hat{\boldsymbol{e}}$, $R_r$ and $\theta \in [0, 2\pi]$ represent the unit vector, receiver CAPA radius, and the azimuth angle within the receiver CAPA, respectively.

Based on the orthogonality between any two basis functions, the orthogonal basis functions $\psi_{l_m}$ satisfy:

$$\int_{S_R} \psi_{l_m}^H \psi_{l_{m'}} ds = \begin{cases} \|\psi_{l_m}\|, l_m = l_{m'}; \\ 0, otherwise. \end{cases} \quad (14)$$

The field radiated by the spatially continuous transmit current can be calculated using a convolution equation and decomposed with appropriate orthogonal basis functions. In the receive region $S_R$, the data is recovered through demodulation using the orthogonal basis functions $\psi_{l_m}$. We then have the input-output representation in terms of $M$ parallel channels:

$$\begin{aligned} y_m &= \int_{S_R} \psi_{l_m}^H \boldsymbol{E}(\boldsymbol{r}) d\boldsymbol{r} = \int_{S_R} \psi_{l_m}^H \int_{S_T} \boldsymbol{G}(\boldsymbol{r},\boldsymbol{s}) \boldsymbol{J}(\boldsymbol{s}) d\boldsymbol{s} d\boldsymbol{r} \\ &= \sum_{n=1}^{N} \int_{S_R} \int_{S_T} \psi_{l_m}^H \boldsymbol{G}(\boldsymbol{r},\boldsymbol{s}) \phi_{l_n} \xi_n d\boldsymbol{s} d\boldsymbol{r}. \end{aligned} \quad (15)$$

We expressed the current of the CAPA transmitter and the electric field at the receiver using OAM mode basis functions. This representation is directly applicable to the design and analysis of practical communication schemes. It accommodates electromagnetic waves generated by arbitrary sources in complex scattering environments and provides a foundation for designing of communication systems based on electromagnetic information theory.

### C. The CAPA Based Electromagnetic Channel

The introduction of transmit and receive basis functions simplifies the modeling of end-to-end CAPA based communication systems, providing a practical approach to analyze the systems with CAPA. By adopting a widely accepted method to define the channel matrix, the analysis becomes more tractable and efficient.

We denote by $h_{mn}$ the coupling coefficient between the transmitter OAM mode $l_m$ and the receiver OAM mode $l_n$. According to Eq. (15), $h_{mn}$ can be given as follows:

$$h_{mn} = \int_{S_R} \int_{S_T} \psi_{l_m}^H G(r,s) \phi_{l_n} ds dr. \quad (16)$$

By introducing the orthogonal basis of OAM modes, we obtain a discrete representation of the CAPA based communication channel. Although this representation may potentially be infinite-dimensional, it provides a foundation for the analysis presented in the subsequent sections.

The MIMO representation of a communication system between two spatially-continuous spaces of arbitrary shape and position can be expressed as follows:

$$y_m = \sum_{n=0}^{N-1} h_{mn} x_n + z_m, \quad (17)$$

where $z_m$ is Gaussian noise with zero-mean and variance of $\sigma^2$, corresponding to the received OAM mode $l_m$.

When the transmitter and receiver CAPAs are approximately arranged with $K$ and $V$ elements, respectively, the classical discrete representation of the MIMO matrix can be further provided as follows:

$$Y = HX + z = R_\psi^H G_{EM} T_\phi X + z = \Lambda X + z, \quad (18)$$

where $z$ is noise vector.

The OAM mode matrices $T_\phi$ at the transmitter and $R_\psi^H$ at the receiver are respectively defined as follows:

$$T_\phi = \begin{bmatrix} \phi_{11} & \cdots & \phi_{1n} & \cdots & \phi_{1N} \\ \vdots & \ddots & \vdots & \ddots & \vdots \\ \phi_{k1} & \cdots & \phi_{kn} & \cdots & \phi_{kN} \\ \vdots & \ddots & \vdots & \ddots & \vdots \\ \phi_{K1} & \cdots & \phi_{Kn} & \cdots & \phi_{KN} \end{bmatrix}, R_\psi^H = \begin{bmatrix} \psi_{11} & \cdots & \psi_{1v} & \cdots & \psi_{1V} \\ \vdots & \ddots & \vdots & \ddots & \vdots \\ \psi_{m1} & \cdots & \psi_{mv} & \cdots & \psi_{mV} \\ \vdots & \ddots & \vdots & \ddots & \vdots \\ \psi_{M1} & \cdots & \psi_{Mv} & \cdots & \psi_{MV} \end{bmatrix}^H. \quad (19)$$

Here, the phase factors are defined as:

$$\phi_{kn} = \frac{e^{i l_n \varphi_k} \hat{j}}{\sqrt{K}}, \psi_{mv} = \frac{e^{i l_m \theta_v} \hat{e}}{\sqrt{V}}, \quad (20)$$

where the azimuth angles are given by $\theta_v = \frac{2\pi v}{V}$ and $\varphi_k = \frac{2\pi k}{K}$.

The $G_{EM}$ is a matrix composed of Green's functions determined by the positions of the transmitter and receiver antenna elements, and is expressed as:

$$G_{EM} = \begin{bmatrix} G(1,1) & \cdots & G(1,k) & \cdots & G(1,K) \\ \vdots & \ddots & \vdots & \ddots & \vdots \\ G(v,1) & \cdots & G(v,k) & \cdots & G(v,K) \\ \vdots & \ddots & \vdots & \ddots & \vdots \\ G(V,1) & \cdots & G(V,k) & \cdots & G(V,K) \end{bmatrix}. \quad (21)$$

The orthogonal basis functions for OAM modes are chosen such that the matrix $\Lambda$ in Eq. (18) becomes diagonal with elements $h_{mm}$, establishing a one-to-one correspondence between the transmitter OAM mode $l_n$ and the receiver OAM mode $l_m$.

### D. The Spectrum Efficiency And Effective Degrees Of Freedom

Assuming that the power is allocated on average to each OAM mode, the total transmit power is denoted as $P_t$. The spectrum efficiency of our proposed CAPA based OAM transmission system can be derived as follows:

$$C = \sum_{m=0}^{M-1} \log_2 \left( 1 + |h_{mm}|^2 SNR \right), \quad (22)$$

where $SNR = \frac{P_t}{M\sigma^2}$ represents the average transmit signal-to-noise ratio (SNR) per OAM mode.

To represent the SE upper bound of a CAPA, we introduce the coupling strength $\gamma_{S_R S_T}$ [16] between the transmitter region $S_T$ and the receiver region $S_R$, which can also be interpreted as the sum of the power transfer coefficients of all effective modes between the two regions. The coupling strength $\gamma_{S_R S_T}$ can be given as follows:

$$\begin{aligned} \gamma_{S_R S_T} &= \sum_{m=0}^{M-1} |h_{mm}|^2 \\ &= \int_{S_R} \int_{S_T} \psi_{l_m}^H G(r,s) \phi_{l_m} \phi_{l_m}^H G^H(r,s) \psi_{l_m} ds dr \\ &= \frac{1}{2\pi R_r} \int_{S_T} K_T(s,s') ds dr = \frac{S_R S_T}{(4\pi d)^2}, \end{aligned} \quad (23)$$

where the kernel function $K_T(s,s')$ is given by

$$K_T(s,s') = K_T(s',s) = \int_{S_R} G^H(r,s) G(r,s') dr. \quad (24)$$

The Eq. (23) indicates that the coupling strength depends solely on the apertures and relative positions of the two antennas. This result is related to the choice of basis functions, as the OAM modes basis functions are complete and orthogonal.

The CAPA based effective degrees of freedom (EDoF) [17] can be expressed as follows:

$$\varepsilon = \frac{\left( \int_{S_R} \int_{S_T} G(r,s) G^H(r,s) ds dr \right)^2}{\int_{S_T} \int_{S_T} K_T(s,s') K_T^H(s,s') ds ds'}. \quad (25)$$

When the number of orthogonal basis functions exceeds a certain saturation point, the eigenvalues corresponding to the exceeded basis functions will become too small to significantly contribute to the EDoF, resulting in the EDoF reaching its maximum value.

The upper bound of the SE for the proposed CAPA based OAM orthogonal transmission can be derived as follows:

$$C = \log_2\left[\prod_{m=0}^{M-1}(1+|h_{mm}|^2 SNR)\right]$$
$$\leq M\log_2 \frac{1}{M}\left[\sum_{m=0}^{M-1}(1+|h_{mm}|^2 SNR)\right] = M\log_2\left(1+\frac{SNR}{M}\sum_{m=0}^{M-1}|h_{mm}|^2\right)$$
$$\leq \varepsilon \log_2\left(1+\frac{SNR}{\varepsilon}\sum_{m=0}^{\varepsilon-1}|h_{mm}|^2\right) = \varepsilon \log_2\left(1+\frac{SNR\gamma_{S_R S_T}}{\varepsilon}\right). \quad (26)$$

## IV. PERFORMANCE EVALUATIONS

In this section, numerical simulation results are presented to evaluate the performance for our proposed CAPA based OAM orthogonal transmission scheme. We focus on the line-of-sight (LOS) scenario while the transmitter and receiver CAPAs are aligned with each other. The communication system operates in the 5.8 GHz frequency band, with the radii of both the transmitter and receiver CAPAs set to $20\lambda$ unless specified otherwise.

Figure 2 shows the spectrum efficiencies of CAPA-based and UCA-based OAM transmission under different SNRs. As shown in Figure 2, the SE of the UCA with an element spacing of $\lambda/2$ is higher than that of UCAs with 8 and 4 elements. The SE of CAPA based OAM transmission with 8 OAM modes is greater than that of transmission with 4 OAM modes. Theoretically, a CAPA can transmit an arbitrary number of mutually orthogonal OAM modes, achieving the upper bound of SE. From an electromagnetic perspective, this is because the CAPA fully exploits the spatial properties of electromagnetic waves, thereby maximizing the information-carrying capacity. From an information-theoretic perspective, the presence of more mutually orthogonal OAM modes enables them to carry more information, which can then be effectively recovered at the receiver.

Figure 3 shows the SE of CAPA-based and UCA-based OAM transmission at different transmission distances. As shown in Figure 3, compared to a discrete UCA, the CAPA based OAM transmission achieves higher SE even over longer distances, thereby providing improved coverage in both the near-field and far-field regions. The SE of OAM transmission with 16 OAM modes is greater than that of transmission with 8 OAM modes. We also observe a steady attenuation of SE as the transmission extends from the near field to the far field. The Green's function-based channel model between the transmitter and receiver CAPAs can more accurately model the electromagnetic channel, enabling CAPA based OAM transmission to achieve higher channel gain at the receiver.

Figure 4 shows the EDoF at 5.8GHz and 24GHz frequency bands under different transmission distances and antenna radii.

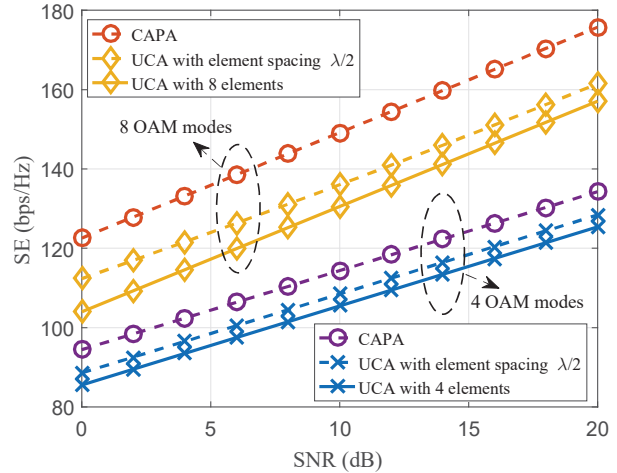

Fig. 2. The spectrum efficiencies of CAPA-based and UCA-based OAM transmission under different SNRs.

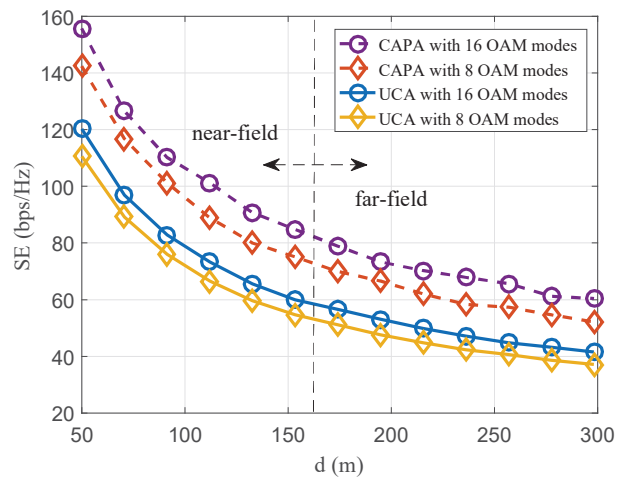

Fig. 3. The spectrum efficiencies of CAPA-based and UCA-based OAM transmission at different transmission distances.

As shown in Figure 4, the EDoF decreases with increasing transmission distance and drops to 1 when the transmission distance exceeds the Rayleigh distance $2D^2/\lambda$. The EDoF increases with the antenna radius, and the near-field region expands accordingly. Additionally, the EDoF at the 24 GHz frequency band is higher than that at 5.8 GHz, which is attributed to the shorter wavelength of the 24 GHz band. Figure 4 highlights the effects of transmission distance, antenna size, and frequency on the EDoF, offering valuable insights for exploring the channel capacity upper bound of CAPA.

Figure 5 shows the EDoF of UCA with different numbers of elements and transmission distances. As the number of elements in the UCA increases, the EDoF correspondingly increases and eventually approaches that of CAPA based OAM transmission. Additionally, the upper bound of EDoF varies with the transmission distance, and the number of elements required to reach these respective limits also changes. Compared to $d=100\lambda$, shorter transmission distances require more antenna elements to reach the upper bound of the EDoF, which is also higher than that at longer distances.

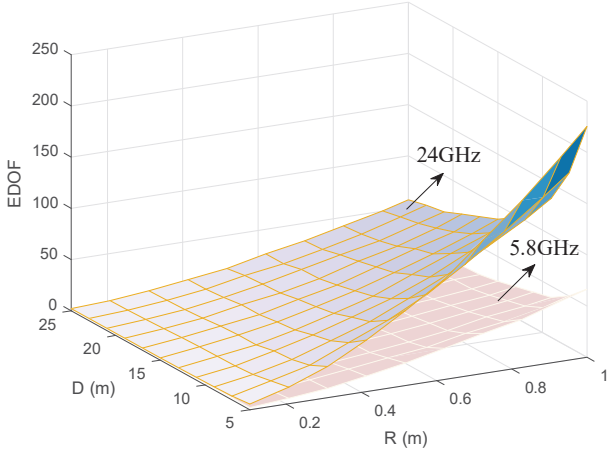

Fig. 4. EDoF at 5.8GHz and 24GHz frequency bands under different transmission distances and antenna radii.

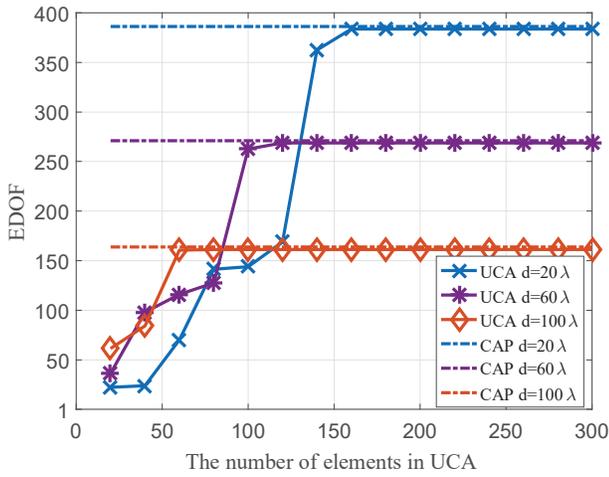

Fig. 5. EDoF of UCA with different numbers of elements and transmission distances.

## V. CONCLUSIONS

In this paper, we proposed a CAPA based OAM orthogonal transmission scheme to realize missive communication. From the perspective of electromagnetic information theory, we designed the surface currents of the CAPA using Fourier basis functions to generate vortex electromagnetic waves for orthogonal transmission. Using the dyadic Greens function, we established a relationship between the surface currents in the CAPA transmitter region and the electric field in the receiver region, thereby deriving the electromagnetic channel for vortex electromagnetic waves. Subsequently, we explored the upper bound of the spectrum efficiency for CAPA based OAM orthogonal transmission. Simulation results demonstrate that the proposed scheme achieved high-capacity communication of vortex electromagnetic waves. This paper establishes the theoretical foundation for applying EIT to the orthogonal transmission of vortex electromagnetic waves, offering a novel solution for achieving CAPA-based efficient and high-capacity communication.